\documentstyle[twocolumn,prb,epsfig,aps]{revtex}
\begin{document}
\draft
\title{Temperature-dependent $H_{c2}$ anisotropy in MgB$_2$ as inferred from 
measurements on polycrystals}
\author{Sergey L. Bud'ko, and Paul C. Canfield}
\address{Ames Laboratory and Department of
Physics and Astronomy, Iowa State University, Ames, Iowa 50011}
\date{\today}
\maketitle
\begin{abstract}
We present data on temperature-dependent anisotropy of the upper 
critical field of MgB$_2$ obtained from the analysis 
of measurements on high purity, low resistivity polycrystals. 
The anisotropy 
decreases in a monotonic fashion with increase of temperature.
\\
\end{abstract}
The recent discovery \cite{jap} of superconductivity with a high critical 
temperature, $T_c \approx 40$ K, in the simple, binary intermetallic 
compound MgB$_2$ evoked intense experimental and theoretical 
studies of the physical properties of this material that resulted 
in understanding of superconductivity in this material as being 
of the BCS type superconductor in which the observed 
value of $T_c$ is a result of the anisotropy of the electron-phonon 
coupling and anharmonicity in the relevant (boron $E_{2g}$) phonon 
modes. \cite{yildirim,choi1,choi2} 

One of the imporant superconducting properties of MgB$_2$ is the 
anisotropy of its upper critical field. Reported values of $\gamma = 
H_{c2}^{ab}/H_{c2}^c$ span between $\gamma \approx 1$ \cite{chen} and 
$\gamma = 9-13$. \cite{shinde} These values were obtained on MgB$_2$ 
in different forms and sample quality. For sub-$mm$ single 
crystallites of 
magnesium diboride the $H_{c2}$ anisotropy was communicated to be in 
the range of $\gamma = 1.7-3$. \cite{kim,mxu,slee,pradhan} Recently 
the anisotropy of the upper critical field of single crystallites of 
MgB$_2$ was studied using torque magnetometry \cite {angst} and it 
was found to be temperature (and applied field) dependent, 
changing monotonically from 
$\gamma \approx 2.8$ at 35 K to $\gamma \approx 6$ at 15 K. It has 
to be mentioned that while these results \cite {angst} have an 
advantage of 
being obtained by {\it direct} measurements on small single crystals, 
the state-of-the-art single crystals 
\cite{kim,mxu,slee,pradhan,angst,solo} have their $T_c$  
lower ($|\Delta T_c| \gtrsim 1$ K) than that of good polycrystalline 
samples \cite {iso,hinks,AH} 
and also have rather moderate vales of residual resistivity 
ratio: $RRR \approx$ 
5-7 for crystallites as opposed to $RRR \approx$ 20 for high 
purity polycrystalline samples.

Temperature-dependent $\gamma$ implies a breakdown of the standard 
anisotropic Ginzburg-Landau theory with a temperature and field 
independent effective mass anisotropy. Temperature-dependent 
anisotropy of $H_{c2}$, $\gamma(T)$, has been observed in a 
number of materials 
\cite {farrell,willi,muto,uher,argonne} and was found to depend on 
the form and purity of the material. Since establishing the intrinsic 
anisotropy of the upper critical field for MgB$_2$ and its temperature 
dependence is of importance for understanding of the properties of this 
material, we will present an alternate evaluation of the $\gamma(T)$ 
behavior in a wide temperature range ($1.8-35$ K) for samples 
with optimal $T_c = 39.2-39.4$ K and high residual resistivity 
ratio ($RRR \gtrsim 20$). The drawback of the approach is that 
the results are inferred from analysis of the measurements on 
polycrystalline material, however this analysis is robust enough to 
reflect the intrinsic anisotropic properties. In a recent 
communication \cite{aniso} we presented anisotropic $H_{c2}$ data 
for $T \gtrsim 25$ K and extracted a value of $\gamma$(25 K)
$ \approx$ 6. In this report we extend these data so as to determine 
the full $\gamma (T)$ plot.

\begin{figure}
\epsfxsize=0.8\hsize
\centerline{
\vbox{
\epsffile{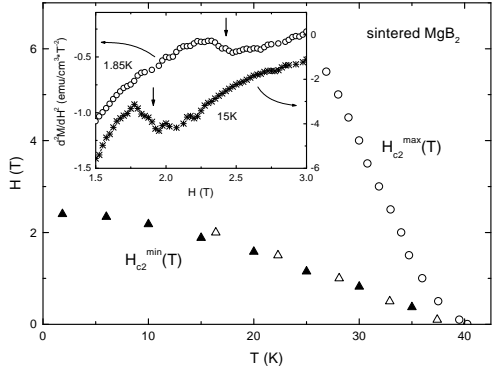}
}}
\vskip \baselineskip
\caption{Anisotropic $H_{c2}(T)$ curves for sintered MgB$_2$. 
Open symbols - from $M(T)|_H$, filled triangles - 
from $M(H)|_T$. 
Inset: examples of features in smoothed $d^2M/dH^2$ curves, 
$H_{c2}^{min}$ are marked with arrows.}
\label{MH}
\end{figure}

Anisotropic $H_{c2}^{min}(T)$ and $H_{c2}^{max}(T)$ data for 
$T \ge 25$ K obtained from the analysis of the 
temperature-dependent magnetization of randomly (continuously) 
oriented MgB$_2$ powders are readily available from Ref. 
\onlinecite{aniso}. 
Applying the qualitative arguments used in Ref. \onlinecite {aniso} 
for $M(T)|_H$ data to magnetization isotherms, 
$M(H)|_T$, \cite {madison} one would 
expect to detect an anomaly at $H_{c2}^{min}$. As in the $M(T)|_H$ 
case the feature should be present for any continuous (but not 
necessary random) distribution of grains. Some theoretical 
discussion, albeit with additional approximations, related to 
the anomaly in second derivative of $M(H)|_T$ was presented  
more than a decade ago \cite {GKL} in relation to high temperature 
copper oxide superconductors. In the case of MgB$_2$  
(sintered sample similar to the one used in [\onlinecite {aniso}]) the 
anomaly in the second 
derivative is clearly seen (see inset to Fig. \ref {MH}). The 
temperature-dependent $H_{c2}^{min}(T)$ data between 1.8 K and 35 K 
was obtained by monitoring this feature at different temperatures 
(see Fig. \ref {MH}). The results deduced from the magnetization 
data taken along different lines in the $H-T$ space are consistent.

\begin{figure}
\epsfxsize=0.8\hsize
\vbox{
\centerline{
\epsffile{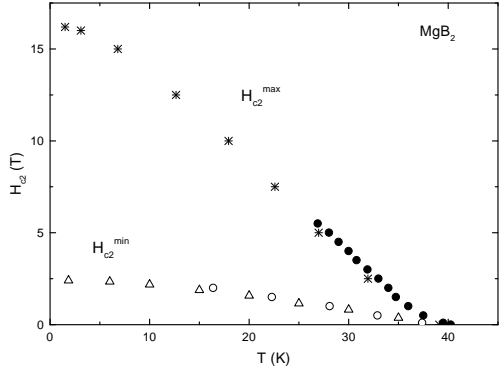}
}}
\caption{Combined anisotropic $H-T$ phase diagram for MgB$_2$. Symbols: 
circles (open and filled) - from $M(T)|_H$ (Ref. 21), 
triangles - from $M(H)|_T$, 
astericks - from polycrystalline $H_{c2}(T)$ (Ref. 25).}
\label{Hc2}
\end{figure}

Upon application of $H \ge H_{c2}^{max}|_T$ all grains in a 
polycrystalline sample become normal, i.e. $H_{c2}^{max}(T)$ coincides 
with $H_{c2}(T)$ measured on a polycrystal. Since the polycrystalline 
$H_{c2}$ is very similar for our sintered pellets \cite {DKF} and wire 
segments \cite {AH,LANL}, we will use the $H_{c2}(T)$ data for wire 
segments \cite {LANL} as an approximation  for $H_{c2}^{max}(T)$ below 
25 K. The data are consistent with the results obtained by analysis 
of $M(T)|_H$ curves \cite {aniso} in the shared temperature region 
(above 25 K). The combined $H-T$ phase diagram for a whole temperature 
range is presented in Fig. \ref{Hc2}. The anisotropy of $H_{c2}$, 
$\gamma (T)$, is 
straightforwardly determined from this phase diagram.

\begin{figure}
\epsfxsize=0.8\hsize
\vbox{
\centerline{
\epsffile{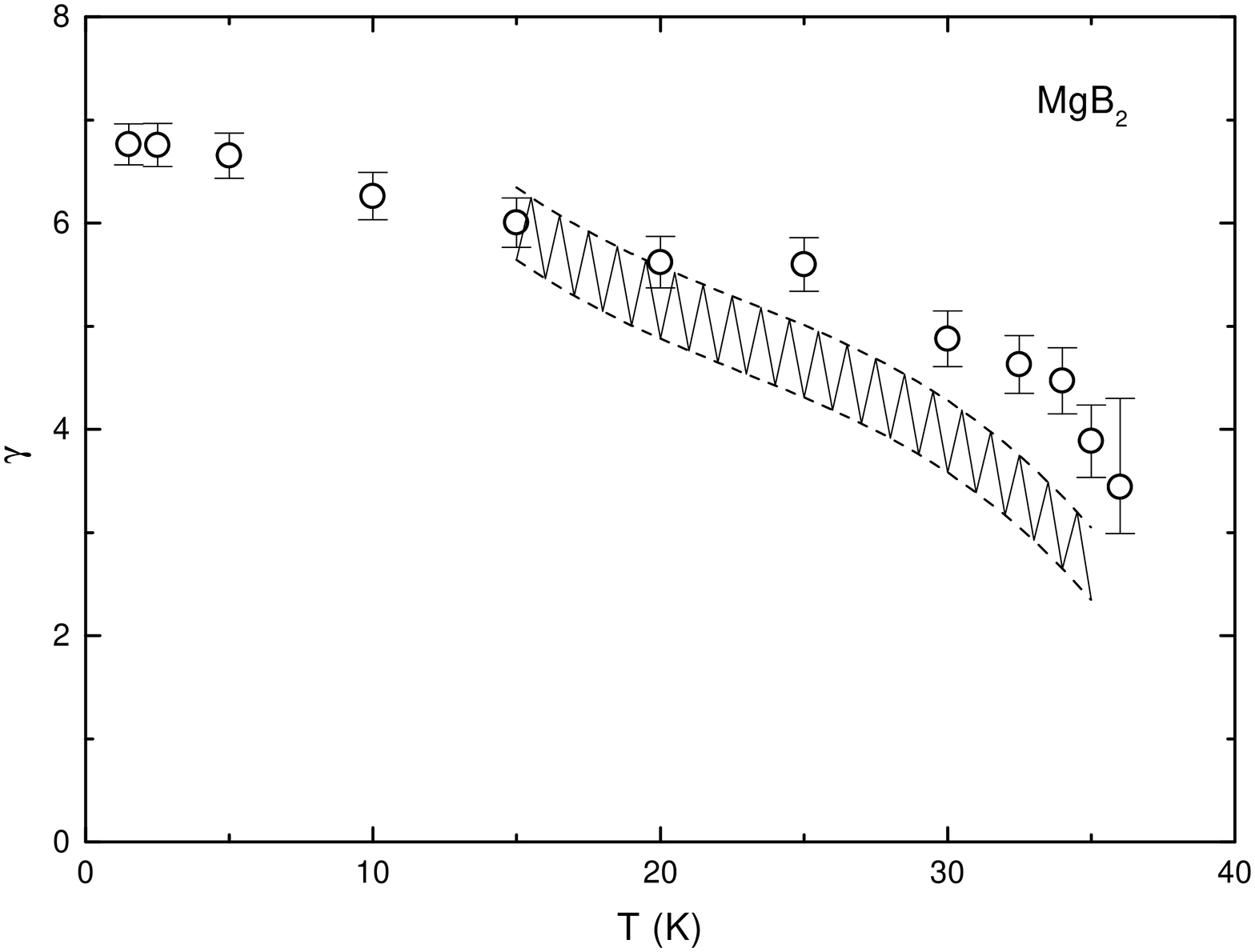}
}}
\caption{Temperature-dependent anisotropy of the upper critical field. 
The range of data from Ref. 11 is shown as a hatched area between 
dashed lines for comparison.}
\label{gammaT}
\end{figure}

Temperature-dependent anisotropy of the upper critical field of 
magnesium diboride inferred from the measurements on  
polycrystalline samples 
is shown in Fig. \ref{gammaT} together with the data from Ref. 
\onlinecite {angst}. 
Our data show a similar, but somewhat less pronounced, temperature 
dependence of the anisotropy: $\gamma$ changes from 3.5 to 7 with 
decrease of the temperature from 36 K down to 1.8 K. 
The fact that the two sets of data are qualitatively similar probably points 
to the intrinsic character of the observed temperature dependence of 
$\gamma (T)$. 
\\

In conclusion, anisotropy of the upper critical field of high purity, 
high $T_c$ 
($T_c \approx 39.2-39.4$ K) and high $RRR$ ($RRR \ge 20$) 
MgB$_2$ samples is 
temperature dependent. 
$\gamma$ decreases monotonically with increase of temperature 
from $\simeq 7$ ($T = 1.8$ K) to $\simeq 4$ ($T = 35$ K). The data 
are qualitatively consistent with the results of the measurements 
on sub-$mm$ single crystals. \cite {angst}
\\

We would like to thank V. G. Kogan for useful discussions.
Ames Laboratory is operated for the
U. S. Department of Energy by Iowa State University under contract No.
W-7405-ENG-82. This work was
supported by the Director of Energy Research, Office of Basic Energy Sciences.

\vfil\eject
\vfil\eject
\end{document}